\documentclass[aps,floats,twocolumn,prb]{revtex4-2}
\usepackage[unicode,linktocpage=true, pdfusetitle]{hyperref}
\hypersetup{colorlinks=true,linkcolor=blue,urlcolor =blue,citecolor=blue,anchorcolor=blue}
\usepackage{amsmath}
\usepackage{graphicx}
\usepackage{xcolor}
\usepackage{amsmath,amssymb,amsfonts, bm}

\begin{document}

\title{Relation between the areal rate of Zitterbewegung and Berry curvature in two-band systems
}

\author{Sonja Predin}
\email{sonja@ipb.ac.rs}
\affiliation{Institute of Physics Belgrade, University of Belgrade, Pregrevica 118, 11080 Belgrade, Serbia}

\begin{abstract}
We introduce the \textit{Zitterbewegung areal rate} operator to study the relation between the 
orientation of trembling motion and band topology. Constructed from the oscillatory interband 
displacement and velocity, this operator distinguishes clockwise from anticlockwise motion.
For a generic gapped two-band Hamiltonian in two dimensions, we show that the Zitterbewegung areal 
rate is proportional to the Berry curvature. 
The derivation uses only spectral projectors and is therefore gauge invariant. We also demonstrate that the Zitterbewegung areal rate operator is time independent and a constant of motion at each momentum, although the displacement and velocity from which it is constructed oscillate in time. 
Its sign distinguishes anticlockwise from clockwise Zitterbewegung and coincides 
with the sign of the Berry curvature. 
For wave packets narrowly localized around the isolated massive Dirac points, the Zitterbewegung 
chirality equals the sign of the local Chern contribution. The Chern number can therefore be 
reconstructed from the orientations of the Zitterbewegung motion around the individual Dirac points.
\end{abstract}

\maketitle

\section{Introduction}
\label{introduction}

The geometry of Bloch states in momentum space is encoded in the
quantum geometric tensor, whose real and imaginary parts determine the
quantum metric and the Berry curvature, respectively. In two dimensions,
the integral of the Berry curvature over the Brillouin zone 
is related to the first Chern number~\cite{tkk,berry}. The Berry curvature
also governs anomalous transport and orbital magnetic
responses~\cite{berry,nagaosa2010,qiang2026}, while the quantum metric
contributes to the superfluid weight of multiband superconductors
\cite{peotta2015} and to the stability of fractional Chern-insulator
states~\cite{parameswaran2012}. Band geometry also manifests itself in
real-time dynamics. In particular, wave-packet motion has been used to
probe the Berry curvature and the topology of Bloch
bands~\cite{Aidelsburger2013,Flaeschner2015}.

Zitterbewegung is an important quantum-dynamical phenomenon that arises 
from interference between different energy bands. It was
first predicted by Schr\"odinger for relativistic electrons governed by
the Dirac equation~\cite{Schroedinger}. Schliemann \textit{et al.} later
showed that an analogous motion in spin-orbit-coupled semiconductors
arises from interband mixing~\cite{Schliemann2005}. Zitterbewegung has
since been studied in graphene~\cite{rusin2008,mireles2022}, carbon
nanotubes~\cite{zawadzki2006}, ultracold atoms~\cite{vaishnav2008}, 
Weyl semimetals~\cite{li2016},
semiconductors~\cite{schliemann2006, shen2022}, topological insulators
\cite{dora2012,shi2013}, and twisted bilayer systems
\cite{lavor2021}.

Zitterbewegung provides a natural setting for studying the relation between interband dynamics and the 
geometry and topology of individual Bloch bands. Previous studies have shown that wave-packet and 
Zitterbewegung dynamics can carry topological information.
In particular, Li \textit{et al.} demonstrated that the center-of-mass motion of a wave packet, 
containing both directed and transient Zitterbewegung contributions, 
can reveal the chirality of a Weyl point~\cite{li2016}.
Moreover, Liu \textit{et al.} experimentally observed valley dependent self-rotation of photonic wave 
packets and related it theoretically to Zitterbewegung oscillations~\cite{buljan2021}.
More directly related to the present work, Shen \textit{et al.} showed that the orientation of 
Zitterbewegung trajectories in spin-$J$ Dirac models depends on the valley index and the sign of the 
mass term, which changes across a band inversion, thereby relating Zitterbewegung to band 
topology~\cite{shen2022}. 

Motivated by these results, we introduce a new quantity, which we call the \textit{areal rate of 
Zitterbewegung}, to characterize Zitterbewegung chirality. 
Since chirality distinguishes clockwise from anticlockwise motion, this quantity should change sign when 
the orientation of the trajectory is reversed. A natural choice is therefore an antisymmetric 
combination of position and velocity. At fixed momentum $\mathbf k$, we denote the field-free interband 
oscillatory part of Heisenberg position operator by $\delta\hat{\mathbf r}(\mathbf k,t)$ and define
\begin{align}
\widehat{\mathcal R}_{\mathrm{ZB}}(t)
&=
\frac{1}{2}
\left[
\delta r_x(t)\,\delta \dot r_y(t)
-
\delta r_y(t)\,\delta \dot r_x(t)
\right].
\end{align}
In this paper, we establish an exact analytical relation between band geometry, 
specifically the Berry curvature, and Zitterbewegung in generic two-band system in
two dimensions 
\begin{align}
\widehat{\mathcal R}_{\mathrm{ZB}}(\mathbf k)
=
\frac{\omega(\mathbf k)}{2}
\Omega_-(\mathbf k)\,\mathbb I,
\end{align}
where $\omega(\mathbf k)=E_+(\mathbf k)-E_-(\mathbf k)>0$ is the
frequency of the interband Zitterbewegung motion, $\Omega_-(\mathbf k)$ is the Berry
curvature of the lower band, and we set $\hbar=1$. 
The derivation is written entirely in terms of spectral projectors and does not require a particular 
gauge choice for the Bloch eigenstates. Although the position and velocity operators oscillate in time, 
their time dependence cancels in the antisymmetric product. The areal rate operator is therefore time 
independent, and a constant of motion at fixed momentum. Moreover, because it is proportional to the 
identity, its expectation value does not depend on the initial internal two component state.

Furthermore, since $\omega(\mathbf k)>0$, the areal rate has the same sign as the local Berry curvature. 
The chirality of the Zitterbewegung motion is therefore determined by the sign of the Berry curvature, 
consistent with the valley and mass dependent direction of the motion found by Liu 
\textit{et al.} and Shen \textit{et al.}~\cite{buljan2021,shen2022}.
In the two-dimensional two-band models considered here, the same parameters also determine the sign of 
the Berry curvature, suggesting a connection between Zitterbewegung chirality and local band geometry. 
In this paper, we make this connection explicit by introducing the \textit{areal rate of 
Zitterbewegung}, a gauge-invariant signed quantity that characterizes the orientation of the interband 
motion. 

Since Zitterbewegung center-of-mass dynamics is already experimentally accessible in photonic and 
ultracold atom systems, this reconstruction could be implemented by resolving the orientation of the 
motion near the individual Dirac points~\cite{lovett2023,leblanc2013,qu2013}.

The paper is organized as follows. In Sec.~\ref{formalism}, we formulate the interband Zitterbewegung 
dynamics using spectral projectors. In Sec.~\ref{arealZb}, we introduce the \textit{areal rate of 
Zitterbewegung}, derive its exact relation to the Berry curvature for a generic two-band Hamiltonian, 
and extend the result to wave packets. We then relate the Zitterbewegung chirality near a massive Dirac 
point to its contribution to the Chern number and illustrate the result using a two-dimensional Chern
insulator model. We close with a summary and an outlook in Sec.~\ref{conclusions}. Technical details of 
the derivation are given in Appendix~\ref{app:areal_rate}.

\section{Zitterbewegung in projector form}
\label{formalism}

In this paper, we mainly consider a generic two-band Hamiltonian
$\mathcal{H}(\mathbf{k})$. However, it
is useful to begin with the general spectral-projector representation.
We impose the two-band restriction explicitly below, at the point where
it becomes necessary. Throughout the paper, we set $\hbar\equiv 1$.

In the Heisenberg picture, the position operator for generic Hamiltonian
$ \mathcal{H}(\mathbf{k}) $ evolves as \cite{david2010}
\begin{align}
\hat{\mathbf r}(t)=e^{i\mathcal H t}\,\hat{\mathbf r}(0)\, e^{-i\mathcal H t}.
\end{align}

We use the spectral (projector) decomposition
\begin{align}
\mathcal H(\mathbf k)= \sum_{a} E_a(\mathbf k)\, Q_a(\mathbf k),
\end{align}
where the projectors $Q_a$ are Hermitian, mutually orthogonal, and complete,
\begin{align}
Q_a Q_b=\delta_{ab}Q_a
\label{orthogonality}
\end{align}
and 
\begin{align}
\sum_a Q_a=\mathbb{I}.
\label{project_sum}
\end{align}

Using this decomposition, the position operator separates into an intraband drift term and an interband 
oscillatory contribution,
\begin{align}
\hat{\mathbf r}(t) = \hat{\mathbf r}(0) + t \sum_a \mathbf{V}_a Q_a + \sum_{a \neq b} \left(e^{i 
\omega_{ab}t}-1\right) \mathbf{Z}_{ab},
\label{position_projector}
\end{align}
where $\mathbf{Z}_{ab}= i Q_a \partial_{\mathbf{k}} Q_b$ encode interband Zitterbewegung oscillations at 
frequencies $\omega_{ab}=E_a-E_b$, and $\mathbf{V}_a=\partial E_a/\partial \mathbf{k}$ denotes the band-
group velocity. Thus, the Zitterbewegung contribution arises from the interband terms $ Q_a 
\partial_{k_i}Q_b $ with $ a \neq b $.


For the analysis below, we consider only the Zitterbewegung contribution and rewrite it as
\begin{align}
\delta \hat r_i(t)
=
\sum_{a\neq b}
e^{i\omega_{ab}t}Z_{ab}^i .
\label{delta_r_general}
\end{align}
The time-independent term $-\sum_{a\neq b}Z_{ab}^i$ in Eq.~\ref{position_projector}
represents a constant displacement of the center of the oscillatory
motion and is neglected together with the remaining nonoscillatory
part of the position operator.

We now restrict the analysis to a two-band Hamiltonian. Denoting the
two bands by $+$ and $-$, Eq.~\ref{delta_r_general} becomes

\begin{align}
\delta\hat{r}_i(t) = & \, i \left(Q_{+}\partial_{k_i}Q_{-} + Q_{-}\partial_{k_i}Q_{+}\right)\cos(\omega t) \nonumber \\
&- \left(Q_{+}\partial_{k_i}Q_{-} - Q_{-}\partial_{k_i}Q_{+}\right)\sin(\omega t),
\label{posCOM}
\end{align}
where $\omega = E_{+}(\bm{k})-E_{-}(\bm{k})$.

Any Hermitian two-band Bloch Hamiltonian can be written as in the terms of Pauli matrices $ \sigma_i $
\begin{align}
\mathcal H(\mathbf k)
=
d_0(\mathbf k)\mathbb I
+
\mathbf d(\mathbf k)\cdot\boldsymbol{\sigma}.
\label{generalDirac}
\end{align}
The scalar term $d_0(\mathbf k)\mathbb I$ affects only the intraband drift, shifts the energy bands 
and leaves the topology unchanged. Since we are
interested in the interband Zitterbewegung contribution, we set $d_0=0$ and write

\begin{align}
\mathcal{H}(\mathbf{k})=\sum_{i=1}^{3} d_i(\mathbf{k})\,\sigma_i ,
\label{HamDirac}
\end{align}
with eigenvalues $E_\pm=\pm d$, where $d=\sqrt{d_1^2+d_2^2+d_3^2}$.
The gap closes at points where $\bm{d}(\mathbf{k})=0$. 

Introducing $\hat{\mathbf d}=\mathbf d/d$, the corresponding spectral
projectors are
\begin{align}
Q_\pm
=
\frac{1}{2}
\left(
\mathbb I
\pm
\hat{\mathbf d}\cdot\boldsymbol{\sigma}
\right).
\label{projectorsDirac}
\end{align}

Substituting these projectors into Eq.~\ref{posCOM}, we obtain
\begin{align}
\delta\hat r_i(t)
={}&
\frac{1}{2}
\left(
\hat{\mathbf d}\times
\partial_{k_i}\hat{\mathbf d}
\right)\cdot\boldsymbol{\sigma}\,
\cos(\omega t)
\nonumber\\
&+
\frac{1}{2}
\left(
\partial_{k_i}\hat{\mathbf d}
\right)\cdot\boldsymbol{\sigma}\,
\sin(\omega t),
\label{zb_dirac}
\end{align}
with $\omega=2d$.
This expression makes explicit that the Zitterbewegung dynamics are
determined by the momentum space texture of the unit vector
$\hat{\mathbf d}$.

The same spectral projectors provide a gauge-invariant representation
of the Berry curvature. For an isolated band $a$, the Berry curvature is
\begin{align}
\Omega_{\alpha\beta}^{(a)}(\mathbf{k})
=
i\,\mathrm{tr}\!\left[
Q_a
\left[
\partial_{k_\alpha}Q_a,
\partial_{k_\beta}Q_a
\right]
\right],
\label{berry_projector}
\end{align}
while the corresponding quantum metric is
\begin{align}
g_{\alpha\beta}^{(a)}(\mathbf{k})
=
\frac{1}{2}\,
\mathrm{tr}\!\left[
\partial_{k_\alpha}Q_a
\partial_{k_\beta}Q_a
\right].
\label{metric_projector}
\end{align}
These expressions are invariant under local gauge transformations of
the Bloch eigenstates because they are written entirely in terms of the
spectral projector $Q_a$, rather than a particular choice of the
corresponding eigenvector~\cite{mitscherling2025}.

For a two-band Hamiltonian in two dimensions in Eq.~(\ref{HamDirac}), substituting
projectors Eq.~(\ref{projectorsDirac})
into Eqs.~\ref{berry_projector} and \ref{metric_projector} for lower band $ - $ gives
\begin{align}
\Omega_{\alpha\beta}^{(-)}(\mathbf k)
=\frac{1}{2}\, \hat{\mathbf d}\cdot
\left(\partial_{k_\alpha}\hat{\mathbf d} \times
\partial_{k_\beta}\hat{\mathbf d} \right),
\label{berry_dirac}
\end{align}
and
\begin{align}
g_{\alpha\beta}^{(-)}(\mathbf k)
=
\frac{1}{4}
\partial_{k_\alpha}\hat{\mathbf d}\cdot
\partial_{k_\beta}\hat{\mathbf d}.
\label{metric_dirac}
\end{align}
Thus, the Zitterbewegung dynamics, the Berry curvature, and the
quantum metric are all determined by the momentum-space texture of
$\hat{\mathbf d}(\mathbf k)$.

\section{Areal rate of Zitterbewegung and Berry curvature}
\label{arealZb}

Recent photonic studies have demonstrated wave-packet motion with a
well defined sense of rotation associated with Zitterbewegung in
symmetry broken honeycomb lattices and a synthetic Weyl
system~\cite{ye2019,buljan2021}. In symmetry broken honeycomb photonic
lattices, Liu \textit{et al.} experimentally observed valley dependent
wave packet self-rotation in the form of spiraling intensity
patterns~\cite{buljan2021}. Their theoretical analysis related this
self-rotation to Zitterbewegung oscillations of the wave-packet center
of mass, with a frequency determined by the spectral gap. In a
theoretical study of a synthetic Weyl system, Ye \textit{et al.} found
helical Zitterbewegung near a Weyl point and related it to the
non-Abelian Berry connection~\cite{ye2019}.

Related theoretical work has connected the orientation of Zitterbewegung motion to band inversion. For 
spin-$J$ models, Shen \textit{et al.} showed that changing the sign of the effective mass term across a 
gap closing reverses the orientation of the Zitterbewegung trajectory~\cite{shen2022}.

Motivated by these results, we introduce a dynamical observable that
characterizes the orientation of interband Zitterbewegung. We show that,
for a generic two-band Hamiltonian in two dimensions, this observable is
directly related to the Berry curvature.

Let $\delta\hat{\mathbf r}(t)$ denote the interband oscillatory part of
the Heisenberg position operator introduced above. In two dimensions, a
combination of this displacement and its velocity gives the quantum
analogue of the area sweep rate. We define the corresponding
time-dependent quantity
\begin{align}
\widehat{\mathcal R}_{\mathrm{ZB}}(t)
&=
\frac{1}{2}
\left[
\delta r_x(t)\,\delta \dot r_y(t)
-
\delta r_y(t)\,\delta \dot r_x(t)
\right].
\label{areal_rate_definition}
\end{align}
We call this quantity the \textit{Zitterbewegung areal rate}.

Its sign distinguishes clockwise from anticlockwise motion, while its
magnitude gives the corresponding areal rate. We consider a two-band
system, where a single interband pair gives the full Zitterbewegung
motion. In this case, we show that the Zitterbewegung areal rate is
directly related to the Berry curvature. The oscillatory coordinate
obeys
\begin{align}
\delta\dot{\hat{\mathbf r}}(t)
=
i\left[\mathcal H,\delta\hat{\mathbf r}(t)\right].
\end{align}

This definition is similar to the intrinsic self-rotation of a
semiclassical Bloch wave packet, described by
$\frac{1}{2}\langle(\hat{\mathbf r}-\mathbf r_c)
\times\hat{\mathbf v}\rangle$ and related to its orbital magnetic
moment~\cite{berry}. There is, however, an important difference. In the
semiclassical approach, the wave packet is constructed within a single
isolated band. In contrast, $\widehat{\mathcal R}_{\mathrm{ZB}}$ is
defined from the exact interband part of the Heisenberg position
operator.

As shown below, for a two-band Hamiltonian,
$\widehat{\mathcal R}_{\mathrm{ZB}}$ is directly proportional to the
Berry curvature, with the interband frequency as the proportionality
factor. The exact identity derived here is specific to gapped two-band Hamiltonians.

Moreover, the operator $\widehat{\mathcal R}_{\mathrm{ZB}}$ is Hermitian,
since $[\delta\hat{x},\delta\dot{\hat{y}}] =
[\delta\hat{y},\delta\dot{\hat{x}}]$, and therefore its expectation value is real.

Substituting Eq.~(\ref{posCOM}) into Eq.~(\ref{areal_rate_definition}) and using the orthogonality of the projectors, we find that all time-dependent terms cancel, leaving 
\begin{align} \widehat{\mathcal R}_{\mathrm{ZB}} &= i\frac{\omega}{2} \left[ Q_a\frac{\partial Q_b}{\partial k_x} \frac{\partial Q_a}{\partial k_y} - Q_a\frac{\partial Q_b}{\partial k_y} \frac{\partial Q_a}{\partial k_x} \right. \nonumber\\ &\qquad\left. - Q_b\frac{\partial Q_a}{\partial k_x} \frac{\partial Q_b}{\partial k_y} + Q_b\frac{\partial Q_a}{\partial k_y} \frac{\partial Q_b}{\partial k_x} \right]. 
\label{areal_rate2} 
\end{align}

For a two-band system, the completeness relation $Q_++Q_-=\mathbb I$ then gives
\begin{align}
\widehat{\mathcal R}_{\mathrm{ZB}}(\mathbf k)
&=
i\frac{\omega}{2}Q_-
\left[
\frac{\partial Q_-}{\partial k_x},
\frac{\partial Q_-}{\partial k_y}
\right]
-i\frac{\omega}{2}Q_+
\left[
\frac{\partial Q_+}{\partial k_x},
\frac{\partial Q_+}{\partial k_y}
\right].
\label{areal_projector_identity}
\end{align}
The derivation is given in Appendix~\ref{app:areal_rate}.

For a two-band Hamiltonian holds $Q_a=|u_a\rangle\langle u_a|$, and therefore, for any operator X, we have
\begin{align}
Q_aXQ_a &=\langle u_a|X|u_a\rangle Q_a= \textrm{Tr}(Q_aX)\,Q_a .
\label{identity_X}
\end{align} 
Setting
\begin{align}
X=i\left[
\frac{\partial Q_a}{\partial k_x},
\frac{\partial Q_a}{\partial k_y}
\right]
\end{align}
and using the definition of the Berry curvature Eq.~(\ref{berry_projector}), we obtain
\begin{align}
iQ_a\left[
\frac{\partial Q_a}{\partial k_x},
\frac{\partial Q_a}{\partial k_y}
\right]Q_a
=
\Omega_a Q_a.
\label{rank_one_identity}
\end{align}

Multiplying Eq.~(\ref{areal_projector_identity}) from the right by
$Q_-$ and $Q_+$, respectively, and adding the resulting expressions,
we obtain
\begin{align}
\widehat{\mathcal R}_{\mathrm{ZB}}(\mathbf k)
=
\frac{\omega(\mathbf k)}{2}
\Omega_-(\mathbf k)\,\mathbb I.
\label{areal_berry_identity}
\end{align}
A derivation of Eq.~(\ref{areal_berry_identity}) is given in Appendix~\ref{app:areal_rate}.

Equation~(\ref{areal_berry_identity}) gives an exact local relation
between interband Zitterbewegung and the Berry curvature. The sign of
the Zitterbewegung areal rate is fixed by the sign of the Berry
curvature and determines the orientation of the motion.

Cserti and D\'avid~\cite{cserti2010} expressed the Berry curvature in terms of the amplitudes $Z_{aa}$ 
with equal band indices. Since $\omega_{aa}=0$, these amplitudes do not describe oscillatory motion and 
give no contribution to the Zitterbewegung areal rate. In contrast, the areal rate in 
Eq.~(\ref{areal_berry_identity}) is constructed from the interband amplitudes $Z_{ab}$, with $a\neq b$, 
which enter the position operator through terms oscillating at frequencies $\omega_{ab}=E_a-E_b$.

Another consequence of Eq.~(\ref{areal_berry_identity}) is that the
Zitterbewegung areal rate is a constant of motion. Although the
Zitterbewegung position operator oscillates in time,
$\widehat{\mathcal R}_{\mathrm{ZB}}$ is time independent and
proportional to the identity operator.
It therefore commutes with the
Hamiltonian,
\begin{align}
\left[
\mathcal H(\mathbf k),
\widehat{\mathcal R}_{\mathrm{ZB}}(\mathbf k)
\right]
=0,
\end{align}
and hence
\begin{align}
\frac{d}{dt}
\widehat{\mathcal R}_{\mathrm{ZB}}(\mathbf k)
=
i\left[
\mathcal H(\mathbf k),
\widehat{\mathcal R}_{\mathrm{ZB}}(\mathbf k)
\right]
=0.
\end{align}
Thus, at each momentum $\mathbf k$, the areal rate remains constant even
though the Zitterbewegung position oscillates. This result is specific
to the two-band case.

To connect the momentum-resolved result with wave-packet dynamics, we
consider a Gaussian wave packet centered at a band-inversion point,
which we take as the origin of momentum space, $\mathbf{k}_0=0$:
\begin{align}
\vert\psi(\mathbf r,0)\rangle
=
\frac{1}{\sqrt{\pi}\,\Delta}
\exp\left(-\frac{x^2+y^2}{2\Delta^2}\right)
\vert\Phi\rangle.
\label{gaussian}
\end{align}
Here, $\Delta$ is the real-space width of the wave packet and
$\vert\Phi\rangle$ is its normalized initial pseudospin state. The
corresponding normalized momentum-space distribution is
\begin{align}
G_{\mathbf k_0}(\mathbf k)
=
\frac{\Delta^2}{\pi}
\exp\left(-\Delta^2 k^2\right),
\qquad
\int d^2\mathbf k\,G(\mathbf k)=1.
\label{gaussian_momentum}
\end{align}
Thus, in the large $\Delta$ limit, the momentum distribution becomes
sharply localized around the band-inversion point.
Since $\Delta$ is the real-space width, the corresponding momentum-space width is of order $1/\Delta$. The single-cone approximation is valid when this width is much smaller than the distance to the other Dirac points and the packet remains within the momentum region where the linearized Dirac Hamiltonian applies. This condition is required only for the wave-packet application; the exact local identity in Eq.~(\ref{areal_berry_identity}) is independent of $\Delta$.

We parametrize the pseudospin state as
\begin{align}
\vert \Phi \rangle = a \, \vert 1,0\rangle + b \, \vert 0,1\rangle,
\label{initial_state}
\end{align}
corresponding to the eigenbasis of $\sigma_z$. 
The choice of quantization axis is conventional and does not affect the generality of the results.

For a wave packet centered at $\mathbf k_0$, 
we define the wave-packet averaged Zitterbewegung areal rate as 
\begin{align} 
\mathcal R_{\mathrm{ZB}}^{\mathrm{wp}}(\mathbf k_0) = \int d^2\mathbf k\, G_{\mathbf k_0}(\mathbf k)\, 
\left\langle \widehat{\mathcal R}_{\mathrm{ZB}}(\mathbf k) \right\rangle . 
\end{align} 
Using Eq.~(\ref{areal_berry_identity}), we obtain 
\begin{align} 
\mathcal R_{\mathrm{ZB}}^{\mathrm{wp}}(\mathbf k_0) = \frac{1}{2} \int d^2\mathbf k\, G_{\mathbf k_0}
(\mathbf k)\, \omega(\mathbf k)\, \Omega_-(\mathbf k). 
\label{areal_general} 
\end{align} 
Since $\widehat{\mathcal R}_{\mathrm{ZB}}$ is proportional to the identity matrix, this result does not depend on the initial pseudospin state $\lvert\Phi\rangle$.

For the two-band Hamiltonian introduced in
Eq.~(\ref{generalDirac}), the areal rate can also be written as
\begin{align}
\mathcal R_{\mathrm{ZB}}^{\mathrm{wp}}
=
\frac{1}{4}
\int d^2\mathbf k\,
G_{\mathbf k_0}(\mathbf k)\,
\omega(\mathbf k)\,
\hat{\mathbf d}\cdot
\left(
\partial_{k_x}\hat{\mathbf d}
\times
\partial_{k_y}\hat{\mathbf d}
\right).
\label{areal_d_vector}
\end{align}

For the same two-band system, the Berry curvature and the quantum metric
satisfy
\begin{align}
\det g(\mathbf k)
=
\frac{1}{4}\Omega_-^2(\mathbf k).
\label{metric_curvature_relation}
\end{align}
Using this relation, the wave-packet areal rate can also be written as 
\begin{align}
\mathcal R_{\mathrm{ZB}}^{\mathrm{wp}}
=
\int d^2\mathbf k\,
G_{\mathbf k_0}(\mathbf k)\,
\omega(\mathbf k)\,
\textrm{sgn}\!\left[\Omega_-(\mathbf k)\right]
\sqrt{\det g(\mathbf k)}.
\end{align}
Thus, the magnitude of the local areal rate depends on the quantum metric, while the Berry curvature 
determines the direction of the Zitterbewegung motion.
This equality is specific to two-band Hamiltonians in two dimensions and follows from the Pauli-matrix 
algebra, or equivalently from the mapping of the Bloch states to the Bloch sphere~\cite{mera2022}. It 
does not generally hold as an equality in multiband systems.

For a two-band Hamiltonian each Dirac point $D$ contributes to the Chern number. 
The contribution is determined by the chirality $\eta_D$ of the
Dirac point and the sign of its mass $m_D$~\cite{sticlet2012}:
\begin{align}
\mathrm{Ch}_{-}
&=
\sum_D \mathrm{Ch}_D,
&
\mathrm{Ch}_D
&=
\frac{1}{2}\eta_D\textrm{sgn}(m_D).
\label{chern_dirac_sum}
\end{align}
Here,
\begin{align}
m_D&=d_3(D),\\
\eta_D&=\textrm{sgn}(J_D),
&
J_D
&=
\left.
\left[
\partial_{k_x}\mathbf d
\times
\partial_{k_y}\mathbf d
\right]_3
\right|_{\mathbf k=\mathbf D}.
\end{align}
 Moreover,
\begin{align}
\textrm{sgn}\!\left[\Omega_-(D)\right]
=
\eta_D\textrm{sgn}(m_D)
=
\textrm{sgn}(\mathrm{Ch}_D).
\label{curvature_dirac_sign}
\end{align}

For a wave packet narrowly centered at $\mathbf D$ and sampling Berry
curvature of a single sign, we define
\begin{align}
\chi_{\mathrm{ZB}}^D
\equiv
\textrm{sgn}
\left(
\mathcal R_{\mathrm{ZB}}^{\mathrm{wp},D}
\right)
=
\textrm{sgn}(\mathrm{Ch}_D).
\end{align}
Here, $\chi_{\mathrm{ZB}}^D$ denotes the Zitterbewegung chirality and specifies the orientation of 
the motion. The values $\chi_{\mathrm{ZB}}^D=+1$ and
$-1$ correspond to anticlockwise and clockwise motion, respectively.
The Chern number can therefore be reconstructed from the orientation of
the Zitterbewegung motion around the individual Dirac points:
\begin{align}
\mathrm{Ch}_{-}
=
\frac{1}{2}
\sum_D\chi_{\mathrm{ZB}}^D.
\label{chern_from_zb}
\end{align}
The Chern-number reconstruction in Eq.~\eqref{chern_from_zb} is consistent with the relation between local Zitterbewegung directions and topological invariants found by Shen \textit{et al.} for spin-$J$ Dirac models~\cite{shen2022}.

For a fixed Dirac-point chirality, a band inversion changes the sign of
the mass, thereby reversing both the Berry curvature and the direction
of the Zitterbewegung rotation. The Zitterbewegung areal rate thus
provides a direct dynamical signature of the underlying band topology.
For a sufficiently narrow wave packet with nonzero interband coherence, the areal rate of the center-of-
mass trajectory has the same sign as the expectation value of the Zitterbewegung areal-rate operator, 
although the two quantities generally differ in magnitude.

Zitterbewegung has already been observed in photonic systems and ultracold atomic gases by tracking the 
wave-packet center-of-mass~\cite{lovett2023,leblanc2013,qu2013}. For an interband-coherent wave packet 
prepared near a massive Dirac point, measuring both components of this motion would show whether the 
trajectory is clockwise or anticlockwise. Its orientation would determine $\chi_{\mathrm{ZB}}^D$ and 
directly test the relation between Zitterbewegung chirality and the local Berry curvature.

\subsection{Chern-insulator model}
\label{sec:chern_insulator}

We finally illustrate our result using a two-dimensional Chern insulator model on
a square lattice. The model supports both trivial and Chern-insulating
phases. Setting the lattice constant to unity, the Bloch Hamiltonian reads
\begin{align}
\mathcal H(\bm k)
&=
A\left(
\sin k_x\,\sigma_x+\sin k_y\,\sigma_y
\right)
+m(\bm k)\sigma_z,
\label{chern_model}\\
m(\bm k)
&=
M-2B\left(2-\cos k_x-\cos k_y\right),
\end{align}
where $A=\pm 1$, $B$, and $M$ are real model parameters. The corresponding
band energies are
\begin{align}
E_\pm(\bm k)=\pm d(\bm k),
\end{align}
where
\begin{align}
d(\bm k)
=
\sqrt{
A^2\left(\sin^2 k_x+\sin^2 k_y\right)
+m^2(\bm k)
}.
\label{chern_dispersion}
\end{align}

Moreover, the model exhibits four Dirac points $ D \in \left\langle\Gamma, \, X, \,Y, \, \textrm{M} \right\rangle $
\begin{align}
\Gamma&=(0,0), &
X&=(\pi,0),
\nonumber\\
Y&=(0,\pi), &
\mathrm M&=(\pi,\pi).
\end{align}

The low-energy Hamiltonian near each of these points $ \mathbf{k} = D + \textbf{q} $
has the following form
\begin{align}
\mathcal H_D(\bm q)
=
A\left(v_x^D q_x\,\sigma_x+v_y^D q_y\,\sigma_y\right)
+m_D\sigma_z
\label{local_dirac_model}
\end{align}
where $ m_D $ are corresponding mass terms
\begin{align}
m_\Gamma=M,\quad
m_X=m_Y=M-4B,\quad
m_{\mathrm{M}}=M-8B.
\label{dirac_masses}
\end{align}
Furthermore, the chirality of each Dirac cone is determined by
\begin{align}
\eta_D
\equiv
\textrm{sgn}
\left(v_x^D v_y^D\right),
\end{align}
where $ v_x^D=\cos(D_x) $ and $ v_y^D=\cos(D_y) $. 
Therefore
\begin{align}
\eta_\Gamma=\eta_{\mathrm M}=+1,
\qquad
\eta_X=\eta_Y=-1.
\label{dirac_chiralities}
\end{align}

In this case, the dispersion relation becomes
\begin{align}
E_\pm^D(\bm q)
&=
\pm \sqrt{m_D^2+A^2q^2},
\label{dispersion}
\end{align}
so that the interband frequency is
$\omega_D(\bm q)=2\sqrt{m_D^2+A^2q^2}$.

The low-energy Berry curvature for lower band in the vicinity of \(D\) is
\begin{align}
\Omega_-^D(\bm q)
=
\frac{\eta_D A^2m_D}
{2\left(m_D^2+A^2q^2\right)^{3/2}}.
\label{low_berry_curvature}
\end{align}

The corresponding Dirac-cone contribution to the Chern number is
\begin{align}
\mathrm{Ch}_D
&\equiv
\frac{1}{2\pi}
\int_{\mathbb R^2}d^2\bm q\,
\Omega_-^D(\bm q)
\nonumber\\
&=
\frac{\eta_D}{2}\textrm{sgn}(m_D).
\label{local_chern_contribution}
\end{align}

Thus, $\mathrm{Ch}_D$ represents the half-integer contribution of an
individual continuum Dirac cone, whereas only the sum of all
Dirac-cone contributions gives the integer Chern number of the lattice
model.

Substituting the low-energy dispersion relation,
Eq.~\ref{dispersion}, and the Berry curvature near the Dirac point
\(D\), Eq.~\ref{low_berry_curvature}, into the general relation
\ref{areal_berry_identity}, we obtain the Zitterbewegung areal rate operator
near \(D\):
\begin{align}
\widehat{\mathcal R}_{\mathrm{ZB}}^D(\bm q)
&=
\frac{\omega_D(\bm q)}{2}
\Omega_-^D(\bm q)\,\mathbb I
\nonumber\\
&=
\frac{\eta_D A^2m_D}
{2\left(m_D^2+A^2q^2\right)}
\,\mathbb I.
\label{local_areal_rate}
\end{align}
Consequently, for a normalized momentum distribution
\(G_D(\bm q)\) localized around \(D\), we obtain
\begin{align}
\mathcal R_{\mathrm{ZB}}^{\mathrm{wp},D}
\simeq
\int d^2\bm q\,
G_D(\bm q)\,
\frac{\eta_D A^2m_D}
{2\left(m_D^2+A^2q^2\right)}.
\label{wavepacket_local_areal_rate}
\end{align}

Finally, the local Zitterbewegung chirality near the Dirac points has the following form 
\begin{align}
\chi_{\mathrm{ZB}}^D
&\equiv
\textrm{sgn}
\left(
\mathcal R_{\mathrm{ZB}}^{\mathrm{wp},D}
\right)
\nonumber\\
&=
\eta_D\textrm{sgn}(m_D)
=
\textrm{sgn}(\mathrm{Ch}_D)
=
2\,\mathrm{Ch}_D.
\label{local_zb_chirality}
\end{align}
At $m_D=0$, the gap closes and the Zitterbewegung chirality is undefined.

\begin{figure}
\centering
\includegraphics[width=0.45\textwidth]{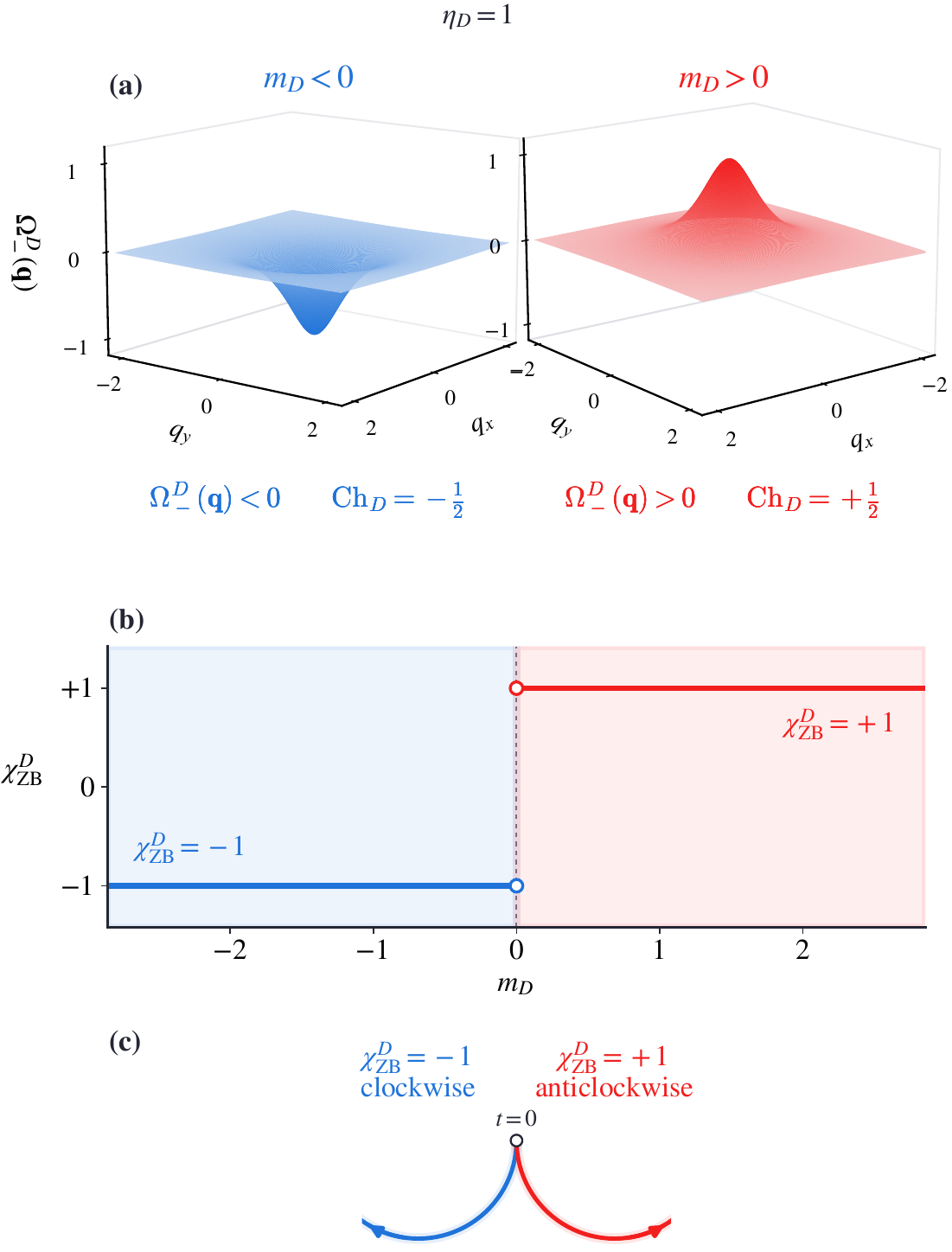}
\caption{
Relation between the Dirac mass, Berry curvature, and Zitterbewegung
chirality near a Dirac point.
For a fixed Dirac-point chirality $\eta_D=+1$, the sign of the Dirac
mass $m_D$ determines the signs of the Berry curvature
$\Omega_-^D$ and the local Chern contribution $\mathrm{Ch}_D$.
(a) Berry curvature near the Dirac point for $m_D<0$ and $m_D>0$.
(b) The Zitterbewegung chirality
$\chi_{\mathrm{ZB}}^D=\textrm{sgn}(\mathrm{Ch}_D)$ changes sign
with $m_D$ and is undefined at $m_D=0$, where the gap closes.
(c) Schematic Zitterbewegung trajectories illustrating the corresponding
change from clockwise to anticlockwise motion.
}
\label{fig1}
\end{figure}

Figure~\ref{fig1} illustrates that, for fixed Dirac-point
chirality $\eta_D=1$, a band inversion reverses the sign of the Berry curvature near
the Dirac point and, consequently, the orientation of the Zitterbewegung
motion. This orientation is determined by
\(\chi_{\mathrm{ZB}}^D=\textrm{sgn}
(\mathcal R_{\mathrm{ZB}}^{\mathrm{wp},D})\), with
\(\chi_{\mathrm{ZB}}^D=+1\) (\(-1\)) corresponding to anticlockwise
(clockwise) motion.

For a Dirac cone with $\eta_D=-1$, the signs of
$\chi_{\mathrm{ZB}}^D$ and the corresponding orientations shown in
Fig.~\ref{fig1} are reversed.

The Chern number of the lower band is obtained by summing the four Dirac cone contributions
\begin{align}
\mathrm{Ch}_-
&=
\frac{1}{2\pi}
\int_{\mathrm{BZ}}
d^2\bm k\,
\Omega_-(\bm k)
\nonumber\\
&=
\frac{1}{2}
\sum_D\chi_{\mathrm{ZB}}^D
\nonumber\\
&=
\frac{1}{2}
\left[
\textrm{sgn}(M)
-2\textrm{sgn}(M-4B)
+\textrm{sgn}(M-8B)
\right].
\label{chern_reconstructed_model}
\end{align}
For $B>0$, the resulting Chern numbers and local Zitterbewegung
chiralities are summarized in Table~\ref{tab:chern}.

The results explicitly demonstrate that the Zitterbewegung chirality at
each Dirac point directly reflects its local topological charge, while the
sum of the individual contributions reconstructs the total Chern number.

\begin{center}
\begin{table}
\begin{tabular}{||l|c|c|c|c||}
\hline\hline
$M$ & $M <0 $ & $0<M<4\,B$ & $4\,B<M<8\,B$ & $M>8\,B$\\
\hline
Ch & 0 & 1 & -1 & 0 \\
\hline
$ \chi_{\mathrm{ZB}}^{\Gamma} $ & $-$ & $+$ & $+$ & $+$\\
$ \chi_{\mathrm{ZB}}^{X} $ & $+$ & $+$ & $-$ & $-$ \\
$ \chi_{\mathrm{ZB}}^{Y} $ & $+$ & $+$ & $-$ & $-$ \\
$ \chi_{\mathrm{ZB}}^{\textrm{M}} $ & $-$ & $-$ & $-$ & $+$ \\
\hline\hline
\end{tabular}
\caption{Chern number and Zitterbewegung chirality at the Dirac points for $B>0$.}
\label{tab:chern}
\end{table}
\end{center}

\section{Conclusions}
\label{conclusions}

We have introduced the Zitterbewegung areal rate to characterize the orientation of Zitterbewegung 
motion. For a generic gapped two-dimensional two-band Hamiltonian, we derived an exact relation between 
the areal rate and the Berry curvature.
We have demonstrated that the Zitterbewegung areal rate operator is time independent, although the 
displacement and velocity from which it is constructed oscillate in time.
The resulting operator is proportional to the identity and is a constant of motion at each momentum, 
independent of the initial internal state.

The sign of the areal rate coincides with that of the lower-band Berry curvature and distinguishes 
anticlockwise from clockwise Zitterbewegung. 
For wave packets narrowly localized around isolated massive 
Dirac points, the Zitterbewegung chirality therefore reproduces the sign of the corresponding local 
Chern contribution. Consequently, the Chern number can be reconstructed from the orientations of the 
Zitterbewegung motion around the individual Dirac points in the two-dimensional two-band system. 

The exact operator identity derived in this work applies to gapped two-band Hamiltonians in two dimensions. The multiband case requires a separate analysis and is outside the scope of the present work.

The relation derived here could be tested in photonic or ultracold atom systems by preparing 
an interband coherent wave packet near a massive Dirac point and determining whether its motion is 
clockwise or anticlockwise.

\emph{Acknowledgments.} The author acknowledges funding provided by the Institute of Physics Belgrade 
through a grant from the Ministry of Science, Technological Development and Innovation of the Republic 
of Serbia. I am grateful to F. Mireles for helpful correspondence and comments on an 
earlier version of the manuscript.

\appendix
\section{Derivation of the Zitterbewegung areal-rate identity}
\label{app:areal_rate}

In this Appendix, we derive Eqs.~(\ref{areal_projector_identity}) and
(\ref{areal_berry_identity}) from the definition of the
Zitterbewegung areal rate in Eq.~(\ref{areal_rate_definition}).
For compactness, we introduce the shorthand notation
\begin{align}
A_i&=Q_a\partial_{k_i}Q_b,
&
B_i&=Q_b\partial_{k_i}Q_a,
\end{align}
where the band labels are chosen such that
$\omega=E_a-E_b>0$. The oscillatory position operator in
Eq.~(\ref{posCOM}) can then be written as
\begin{align}
\delta\hat r_i(t)
={}&
i(A_i+B_i)\cos(\omega t)
-(A_i-B_i)\sin(\omega t).
\label{app:pair_position}
\end{align}

Substituting Eq.~(\ref{app:pair_position}) and its time derivative into
Eq.~(\ref{areal_rate_definition}), we obtain
\begin{align}
\widehat{\mathcal{R}}_{\mathrm {ZB}} & = \frac{\omega}{2} \left[A_x\,A_y - A_y\,A_x + B_x\,B_y - B_y\,B_x\right]\sin(2 \omega t) \nonumber \\
& + i \frac{\omega}{2} \left[A_x\,B_y - A_y\,B_x - B_x\,A_y + B_y\,A_x\right] \nonumber \\
& - i \frac{\omega}{2} \left[A_x\,A_y - A_y\,A_x - B_x\,B_y + B_y\,B_x\right] \cos(2\omega t).
\label{app:areal_rate_ab}
\end{align}

Differentiating the orthogonality relation $Q_aQ_b=0$ for $a\neq b$ gives
\begin{align}
\frac{\partial Q_a}{\partial k_i}\,Q_b = - Q_a \frac{\partial Q_b}{\partial k_i}
\label{app:orthogonality_derivative}
\end{align}
and differentiating the projector identity $Q_a^2=Q_a$ gives
\begin{align}
\frac{\partial Q_a}{\partial k_i}\,Q_a = \frac{\partial Q_a}{\partial k_i} - Q_a \frac{\partial Q_a}{\partial k_i}.
\end{align}

These relations imply
\begin{align}
A_iA_j &=-Q_aQ_b \frac{\partial Q_a}{\partial k_i}\frac{\partial Q_b}{\partial k_j} \\
B_iB_j &= -Q_bQ_a \frac{\partial Q_b}{\partial k_i}\frac{\partial Q_a}{\partial k_j}, \\
A_iB_j &= Q_a\frac{\partial Q_b}{\partial k_i}Q_b\frac{\partial Q_a}{\partial k_j}, \\
B_iA_j &= Q_b\frac{\partial Q_a}{\partial k_i}Q_a\frac{\partial Q_b}{\partial k_j}.
\end{align}
Using the orthogonality relation $Q_aQ_b=0$ once again, we obtain
\begin{align}
A_iA_j&=0,
\label{app:identity_AA}
\\
B_iB_j&=0,
\label{app:identity_BB}
\\
A_iB_j&=Q_a\frac{\partial Q_b}{\partial k_i}\frac{\partial Q_a}{\partial k_j},
\label{app:identity_AB}
\\
B_iA_j&=Q_b\frac{\partial Q_a}{\partial k_i}\frac{\partial Q_b}{\partial k_j}.
\label{app:identity_BA}
\end{align}

Consequently, all terms proportional to $\sin(2\omega t)$
and $\cos(2\omega t)$ in Eq.~(\ref{app:areal_rate_ab})
vanish, leaving
\begin{align}
\widehat{\mathcal{R}}_{\mathrm{ZB}} & = i \frac{\omega}{2} \left(Q_a \frac{\partial Q_b}{\partial k_x} \frac{\partial Q_a}{\partial k_y} - Q_a \frac{\partial Q_b}{\partial k_y} \frac{\partial Q_a}{\partial k_x} \right. \nonumber \\
&  \left. - Q_b \frac{\partial Q_a}{\partial k_x} \frac{\partial Q_b}{\partial k_y}
+  Q_b \frac{\partial Q_a}{\partial k_y} \frac{\partial Q_b}{\partial k_x} \right).  
\label{app:pair_resolved_rate}
\end{align}

Thus, the areal rate associated with a fixed interband pair is time-independent.

For a two-band system, where $a=+$ and $b=-$, this is the full Zitterbewegung
contribution. Using $Q_++Q_-=\mathbb I$, and hence
$\partial_{k_i}Q_+=-\partial_{k_i}Q_-$,
Eq.~(\ref{app:pair_resolved_rate}) reduces to
Eq.~(\ref{areal_projector_identity}) of the main text
\begin{align}
\widehat{\mathcal R}_{\mathrm{ZB}}(\mathbf k)
&=
i\frac{\omega}{2} Q_-\left[\frac{\partial Q_-}{\partial_{k_x}},\frac{\partial Q_-}{\partial_{k_y}}\right] -i\frac{\omega}{2}
Q_+\left[\frac{\partial Q_+}{\partial_{k_x}},\frac{\partial Q_+}{\partial_{k_y}}\right].
\end{align}

Moreover, for a two-band Hamiltonian holds 
\begin{align}
Q_a=|u_a\rangle\langle u_a|.
\end{align}
Therefore, for any operator $X$,
\begin{align}
Q_aXQ_a
&=
|u_a\rangle\langle u_a|X|u_a\rangle\langle u_a|
\nonumber\\
&=
\langle u_a|X|u_a\rangle Q_a
\nonumber\\
&=
\operatorname{Tr}(Q_aX)\,Q_a.
\label{rank_one_general}
\end{align}

Setting
\begin{align}
X=i\left[
\partial_{k_x}Q_a,\partial_{k_y}Q_a
\right]
\end{align}
we find
\begin{align}
iQ_a
\left[
\partial_{k_x}Q_a,\partial_{k_y}Q_a
\right]
Q_a
=
\Omega_a Q_a.
\end{align}

Furthermore, we multiply Eq.~(\ref{areal_projector_identity}) by $Q_-$ and, after some algebra, obtain
\begin{align}
\widehat{\mathcal R}_{\mathrm{ZB}} Q_-= \omega \Omega_- Q_- - i \frac{\omega}{2}\left[\frac{\partial Q_-}{\partial_{k_x}},\frac{\partial Q_-}{\partial_{k_y}}\right]Q_-.
\label{app:identity_multiply_Q-}
\end{align} 
Analogously, multiplying Eq.~(\ref{areal_projector_identity}) by $Q_+$, we obtain
\begin{align}
\widehat{\mathcal R}_{\mathrm{ZB}} Q_+=- \omega \Omega_+ Q_+ + i \frac{\omega}{2}\left[\frac{\partial Q_+}{\partial_{k_x}},\frac{\partial Q_+}{\partial_{k_y}}\right]Q_+.
\label{app:identity_multiply_Q+}
\end{align}

Since $\partial_{k_i}Q_+=-\partial_{k_i}Q_-$, the two commutators are
equal. Furthermore, the commutator
$[\partial_{k_i}Q_a,\partial_{k_j}Q_a]$ commutes with $Q_a$.
Adding Eqs.~(\ref{app:identity_multiply_Q-}) and (\ref{app:identity_multiply_Q+}) and using
Eq.~(\ref{areal_projector_identity}), we obtain
\begin{align}
\widehat{\mathcal R}_{\mathrm{ZB}}
&=
\omega\Omega_-Q_-
-
\omega\Omega_+Q_+
-
\widehat{\mathcal R}_{\mathrm{ZB}}.
\end{align}
Finally, using $\Omega_+=-\Omega_-$ and $Q_++Q_-=I$, we find
\begin{align}
\widehat{\mathcal R}_{\mathrm{ZB}}(\mathbf k)
=
\frac{\omega(\mathbf k)}{2}\Omega_-(\mathbf k)I.
\end{align}



\end{document}